 \documentclass[preprint,12pt,3p]{elsarticle}

\usepackage{graphicx}
\usepackage{epstopdf}
\usepackage{color}
\usepackage{lscape}

\usepackage{amssymb}
\usepackage{amsmath}
 \usepackage{amsthm}

\journal{Physica B}

\begin{document}
\begin{frontmatter}


%
%
\title{Electronic and magnetic properties of quasi-skutterudite PrCo$_2$Ga$_8$ compound} 
\author{Michael O. Ogunbunmi $^{1}$, 
	Buyisiwe M. Sondezi $^{1*}$, 
	Harikrishnan S. Nair $^{1,2}$,
	Andr\'{e} M. Strydom $^{1}$ }
\cortext[cora]{Corresponding author}
\address{$^1$Highly Correlated Matter Research Group, Physics Department, University of Johannesburg, P. O. Box 524, Auckland Park 2006, South Africa}
\address{$^2$Department of Physics, Colorado State University, 200 West Lake Street, Fort Collins, CO 80523-1875, USA}

\cortext[cor1]{bmsondezi@uj.ac.za}
%
%
\begin{abstract}
	PrCo$_2$Ga$_8$ is an orthorhombic quasi-skutterudite type compound which crystallizes in the CaCo$_2$Al$_8$ structure type, with space group $Pbam$ (No. 55). The Pr$^{3+}$ ion has a site symmetry of $C_s$ which predicts a crystal electric field (CEF) level splitting into 9 singlets for $J$ = 4. However, a phase transition at $T_m$ = 1.28~K is observed in electrical resistivity and specific heat results and is reported in this paper. The electrical resistivity shows an upturn below $T_m$ due to the superzone-gap formation. This transition is tuneable in fields and is suppressed to lower temperatures with applied magnetic fields. The electronic specific heat $C_{p}(T)/T$ increases below $T_m$ and reaches a value of 7.37~J/(mol K$^2$) at 0.4~K. The Sommerfeld coefficient, $\gamma$ extracted from the low temperature analysis of $C_\mathrm{4f}(T)/T$ is  637~mJ/(mol K$^2$) indicating a possible mass enhancement of the quasiparticles. The calculated entropy value of 3.05~J/(mol K) is recovered around $T_m$ exhibiting almost 53\% of Rln2, where R is the universal gas constant. Magnetic susceptibility results obeys the Curie-Weiss law for data above 100 K with an estimated effective magnetic moment, $\mu_\mathrm{eff}$ = 3.37~$\mu_B$/Pr and Weiss temperature, $\theta_p$ = $-$124~K.
\end{abstract}
\end{frontmatter}
\section{Introduction}
\label{intro}
\indent
In recent investigations on PrCo$_2$Al$_8$ and PrFe$_2$Al$_8$ compounds which crystallize in the orthorhombic CaCo$_2$Al$_8$ type structure \cite{tougait2005prco, nair2016magnetic}, a long-range magnetic ordering of the Pr$^{3+}$ moment has been reported at 5~K and 4~K respectively. In this structure,  there are four formula units in a unit cell where the Pr atoms are surrounded by nine Al atoms which form a 
polyhedral structure that resembles cages. In addition, the Pr and Co/Fe atoms form 
a chain-like structure parallel to the $c$-axis. The $Pbam$ 
space group allows for one crystallographic position for the Pr atoms, two for Co/Fe 
and 9 for the Al atoms. A detailed description of the crystal structure is given elsewhere \cite{gladyshevskii1983crystal,pottgen2001new}. The Pr$^{3+}$  ion has a site symmetry of $C_s$ which predicts a CEF level splitting into 9 singlets for $J$ = 4 and thus rule out in principle the occurrence of spontaneous magnetic order. For a local rare-earth symmetry with a singlet ground state prediction, a self-induced moment ordering also known as \textquoteleft bootstrap\textquoteright~process can still be achieved in an instance where there is an admixture of the CEF excited states with the ground state singlet under the condition that $J_{ex}$/$\Delta$  exceeds some threshold value (mostly determined from inelastic neutron scattering measurement) resulting in a pseudo-doublet ground state. Here, $J_{ex}$ and $\Delta$ are the exchange integral and the splitting energy between the ground state and the first excited state respectively \cite{anand2014investigations,wallace1973rare}.\\
In the present study we have investigated the electronic and magnetic properties of the iso-structural $\mathrm{PrCo_2Ga_8}$ compound. We show that the substitution of Al for Ga in this structure produces a strong $cf$ hybridization that results in a non-Fermi liquid behaviour in the ground state. We also show that  Pr moment in PrCo$_2$Ga$_8$ develops a magnetic ordering despite its non-magnetic singlet CEF ground state due to its polarization under favourable exchange interaction.\\
\begin{figure}[!h]
	\centering
	\includegraphics[scale=0.8]{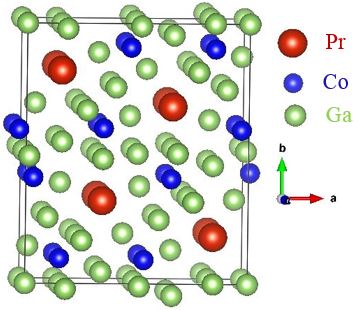}
	\caption{The crystal structure representation of PrCo$_2$Ga$_8$ showing the network of Ga (green) around Pr (red) and Co (blue) atoms viewed down the $c$-axis.\label{structure}}
\end{figure}
\section{Experimental methods}
Polycrystalline samples of $\mathrm{PrCo_2Ga_8}$, $\mathrm{LaCo_2Ga_8}$  and $\mathrm{LaRu_2Ga_8}$ were prepared in an Edmund B{\"u}hler arc melting furnace. Stoichiometric amounts of the high 
purity elements, Pr, Co, Ga, Ru, La and Pr  (4N) were melted on 
the water-cooled Cu-hearth of the arc meltor. The melting was performed in a chamber 
of Ar atmosphere purified using high temperature Zr getter. The resulting ingots were flipped over and remelted to promote homogeneity in the sample. As-cast samples were 
then wrapped in high purity tantalum foil and annealed at 900~$^\circ$C for 14 days in 
an evacuated quartz tube. Room temperature powder X-ray diffraction was recorded using a Rigaku Smartlab diffractometer with Cu-K$\alpha$ radiation.  A Rietveld refinement \cite{rietveld1969profile} carried out results in a straight forward agreement of the powder diffractogram collected and the orthorhombic $Pbam$ space group crystal structure. All the samples were found to be single phase. LaCo$_2$Ga$_8$ was found to show a magnetic ordering of the Co moment below 52~K and is thus not suitable as a non-magnetic reference. We have therefore used LaRu$_2$Ga$_8$ in the place of LaCo$_2$Ga$_8$ after we have confirmed Ru not to be having any local magnetic moment through a magnetic susceptibility measurement. From the Rietveld refinement on the powder sample diffraction pattern, the lattice parameters values obtained are $a$ = 12.2386~\AA, $b$ = 14.3110~\AA ~and $c$ = 4.0560~\AA~  for PrCo$_2$Ga$_8$ which closely agree with a previous report \cite{sichervich1985intermetallic}. The crystal structure of PrCo$_2$Ga$_8$ is shown in Fig.~\ref{structure}. In this structure, the shortest Pr-Pr distance is 4.055~\AA~ which is in agreement with the values found for related iso-structural compounds \cite{tougait2005prco,nair2016magnetic}. The relatively large separation between the Pr atoms suggests they are weakly bonded together which may results in the suppression of the critical temperature in systems with a magnetic order parameter in the ground state. The shortest Pr-Ga separation is 3.096~\AA~(exceeding the sum of their covalent radii by 6\%) and Pr-Co is 3.112\AA~(exceeding the sum of their covalent radii by 11\%). As a consequence, the structure is quasi-skutterudite in nature resembling the filled caged compounds where the Pr atoms are situated in over-sized cages formed by Ga atoms.  For $\mathrm{LaRu_2Ga_8}$, the lattice parameters obtained from Rietveld refinement of the powder pattern are $a$ = 12.6550~\AA,~$b$ = 14.7160~\AA~ and $c$ = 4.1180~\AA. Here, we note that the cell volume of $\mathrm{LaRu_2Ga_8}$ is larger than that of  $\mathrm{PrCo_2Ga_8}$ which is consistent with a previous report \cite{schluter2001ternary}.\\
Magnetic properties were measured using the Magnetic Property Measurement System (by Quantum Design Inc. San Diego) between 1.9~K and 400~K with an external magnetic field up to 7~T. The electrical resistivity measurement down to 0.35~K was carried out using an ETO option of the Physical Property Measurement System (PPMS), also from Quantum Design using the conventional four probe method with contacts made using  a spot welding equipment.  The thermal conductivity, thermopower and electrical resistivity have been measured simultaneously on a sample with the same set of contacts. Specific heat was measured using the quasi-adiabatic thermal relaxation method down to 0.4~K. Both the thermal transport and specific heat were measured on the PPMS. \\
\begin{figure}[!h]
	\centering
	\includegraphics[scale=1.50]{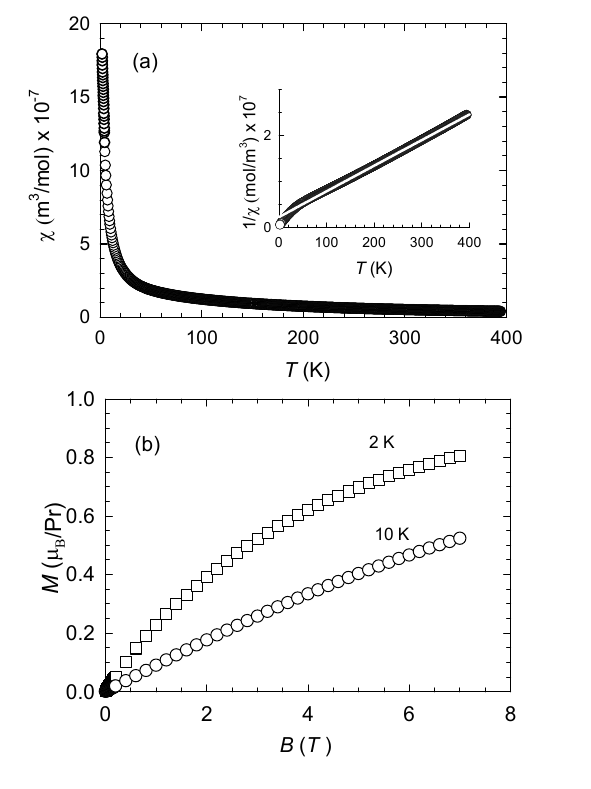}
	\caption{(a) The magnetic susceptibility variation with temperature of  PrCo$_2$Ga$_8$ is shown in the main panel. The inset is a plot of the inverse magnetic susceptibility with a Curie-Weiss fit shown in red solid line. (b) The magnetization of PrCo$_2$Ga$_8$ measured in fields up to 7~T for temperatures 2~K and 10~K is shown .\label{magnetic}}
\end{figure}
\section{Magnetic properties}
The magnetic susceptibility of PrCo$_2$Ga$_8$ measured between 1.9 and 400~K and in an applied field of 0.05~T is presented in Fig. \ref{magnetic}(a). Within this temperature range, no anomaly was observed and the magnetic susceptibility shows a paramagnetic behaviour down to about 20~K before showing a rapid increase at low temperatures with no sign of saturation. This suggests the possibility of a magnetic order parameter at a lower temperature.  The inset is a plot of the inverse susceptibility with a Curie-Weiss fit to the data as indicated by the solid line based on the expression: $\chi^{-1} = 3k_B(T-\theta_p)/N_A\mu^2_\mathrm{eff}$, where k$_B$ and N$_A$ are the Boltzmann's constant and Avogadro's number, respectively. From the least squares fit, the calculated effective magnetic moment, $\mu_\mathrm{eff}$ = 3.37~$\mu_B$/Pr and the paramagnetic Weiss temperature, $\theta_p$ = $-$124~K are obtained. The $\mu_\mathrm{eff}$ value is slightly reduced compared to the value of $g[J(J+1)]^{1/2}$ = 3.58~ $\mu_B$/Pr expected for  a free Pr$^{3+}$ ion. This also suggests that Co does not carry a localized moment in PrCo$_2$Ga$_8$. The negative value of the Weiss temperature suggests a prominent antiferromagnetic (AFM) interactions in the system. The large negative value of the Weiss temperature also gives an indication of a strong $cf$ hybridization in PrCo$_2$Ga$_8$ which influences the nature of the ground state in the compound. The isothermal magnetization presented in Fig.~\ref{magnetic}(b) shows a small curvature at 2~K compared to that of 10~K that is mostly quasi-linear in applied fields. The moment attained at 7~T and 2~K is 0.8~$\mu_B$/f.u which is much reduced compared to the saturation moment of $gJ\mu_B$ = 3.2~$\mu_B$/Pr expected for a free Pr$^{3+}$ ion. This reduction is likely due to the partial quenching of the orbital momentum due to crystal field effects.
  \begin{table}[!b]
  	\centering
  	\setlength{\tabcolsep}{5pt}
  	\caption{\label{tab1}The various $RE$Co$_2$Ga$_8$ ($RE$ = rare-earth elements) compounds reported  along with their critical temperatures for the ones with magnetic phase transition.}
  	\begin{tabular}{llllll}\hline\hline
  		Compound & $T_N$ (K)   & $\theta_p$ (K)   & $\mu_\mathrm{eff}$ ($\mu_B$/f.u) & de Gennes factor   &  Ref.    \\ \hline\hline
  		YbCo$_2$Ga$_8$   & nil   & -  & -    &  0.32 & \cite{fritsch2004antiferromagnetic}    \\
  		EuCo$_2$Ga$_8$   & 20   & $-$2  & 7.81 & 15.75  & \cite{sichevych2009eutm2ga8}   \\
  		CeCo$_2$Ga$_8$ & nil  & -   & 2.45     & 0.18 & \cite{koterlin1989bs,sichervich1985intermetallic,wang2017heavy}    \\
  		PrCo$_2$Ga$_8$ & 1.28  & $-$124   & 3.37 & 0.8  & present report  \\ \hline\hline
  	\end{tabular}
  \end{table}
\section{Specific heat}
\begin{figure}[!t]
	\centering
	\includegraphics[scale=1.2]{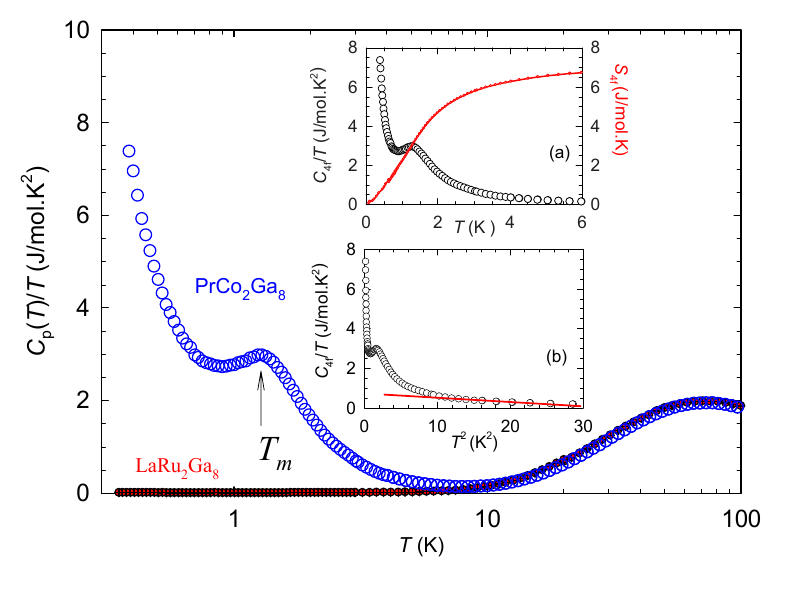}
	\caption{The $C_p(T)/T$ of  PrCo$_2$Ga$_8$ and  LaRu$_2$Ga$_8$ are shown on a semi-log plot in the main panel. $T_m$ is indicated by the arrow. Inset (a) is a plot of temperature variation of $C_{4f}(T)/T$  (black circle) and the entropy, $S_{4f}$ ( red solid line) with axis on the right side. (b) is a plot of $C_{4f}(T)/T$ against $T^2$ with a linear fit.\label{fig1_cp}}
\end{figure}

In Fig.~\ref{fig1_cp}, the specific heat of PrCo$_2$Ga$_8$ plotted as $C_p(T)/T$ together with that of  LaRu$_2$Ga$_8$ to serve as a non-magnetic reference are presented. Below 5~K, $C_p(T)/T$ of  PrCo$_2$Ga$_8$ increases rapidly and terminates in a magnetic phase transition $T_m$ = 1.28~K. We note that this phase transition occurs at a lower temperature compared to other iso-structural compounds where  long-range magnetic transitions have been observed. As mentioned earlier, an AFM ordering in  PrCo$_2$Al$_8$ occurs at 5~K while in  PrFe$_2$Ga$_8$ \cite{Ogunbunmi2017magnetization}, a magnetic ordering believed to be of an AFM origin was observed at a much higher temperature of 17~K which is one order of magnitude higher than the observation in the present case.
For systems with long-range magnetic ordering, the RKKY interaction is expected to decrease as 1/$d^3_{Pr-Pr}$ \cite{nishioka2009novel}, where $d_{Pr-Pr}$ is the nearest distance between two Pr$^{3+}$ ions. 
 In Table \ref{tab1}, we present various $RE$Co$_2$Ga$_8$ compounds reported  along with their critical temperatures and the de Gennes factor, $(g-1)^2J(J+1)$  \cite{jensen1991rare}. From the table, only EuCo$_2$Ga$_8$ and PrCo$_2$Ga$_8$ compounds show magnetic ordering while YbCo$_2$Ga$_8$ and CeCo$_2$Ga$_8$ did not show any magnetic phase transition down to the lowest temperature studied. In view of this, a plot of transition temperature, $T_N$ against the de Gennes factor for EuCo$_2$Ga$_8$ and PrCo$_2$Ga$_8$ will only give a straight line which does not provide us with a clear picture of  how well the $T_N$(K) scales with the de Gennes factor in the present case.

 The magnetic ordering observed in PrCo$_2$Ga$_8$ is however at variance with a singlet ground state prediction based on the $C_s$ site symmetry of Pr$^{3+}$ ion in the crystal structure of this compound. Such observations are largely attributed to induced magnetism which may occur through the formation of a pseudo-doublet ground state arising from the admixture of the ground state singlet and the first CEF level when the magnetic exchange energy is comparable to or higher than the first CEF excitation. In iso-structural compound PrFe$_2$Al$_8$, both macroscopic and microscopic measurements have confirmed the setting in of an AFM ordering below $T_N$ = 4~K \cite{nair2016magnetic,nair2017pr}. Below about 0.8~K, $C_p(T)/T$ shows an upturn that develops into a logarithmic-like divergence. This upturn reaches a value of 7.37~J/mol.K$^2$ at 0.4~K. We have investigated the possible origin of such upturn by considering  the likelyhood of nuclear contribution to the specific heat at low temperatures. Both Pr and Co have only one stable isotope each while Ga has two (Ga-69 and Ga-71 with relative abundances of 60.1\% and 39.9\% respectively). From literature, the low temperature heat capacity of Ga (at 0.3532 K) = 0.0734 mJ/mol.K \cite{seidel1958specific}, Pr ( at 0.3585 K) = 170.9 mJ/mol.K \cite{lounasmaa1964specific} and Co ( at 0.7 K) = 10.2 mJ/mol.K \cite{heer1957hyperfine}. This gives an approximate heat capacity value of about 0.1812 J/mol.K which is about 1-2 order of magnitude lower than the specific heat observed in  PrCo$_2$Ga$_8$ in the same temperature range. These supports the facts that the low temperature upturn does not originate from nuclear effect. The behaviour in PrCo$_2$Ga$_8$ is likely as a result of  proximity to quantum critical point which is a key signature of heavy electron system. Similar observation has been reported recently in iso-structural CeCo$_2$Ga$_8$ \cite{wang2017heavy} albeit the upturn attains only 0.8 J/mol.K$^2$ at about 1~K.\\
In inset (a) of Fig.~\ref{fig1_cp}, the electronic contribution to $C_p(T)$ is plotted as $C_\mathrm{4f}(T)/T$ together with the calculated entropy (red solid line). The entropy has been calculated using the expression: $S_\mathrm{4f}(T') = \int_{0}^{T'}{C_\mathrm{4f}(T)}/{T}$ d$T$. The transition $T_m$ is visible in $C_\mathrm{4f}(T)/T$  and the entropy recovered at $T_m$ is 3.05~J/mol.K which is about 53~\% of Rln(2) expected for a doublet ground state. In inset (b), a plot of $C_\mathrm{4f}(T)/T$ against $T^2$ is plotted. We note that the estimation of $C_\mathrm{4f}(T)$ was only carried out up to about 10~K beyond which the phonon specific heat dominates. Within this temperature range, we observed a plot of  $C_\mathrm{4f}(T)/T$ to be a decreasing function of $T$.  The fit to extract the electronic Sommerfeld coefficient, $\gamma$ by a linear fit to the low temperature region as indicated by the red solid line gives a negative slope and resulting in a value of $\gamma$ = 637~mJ/mol.K$^2$, which suggests a possible quasi-particle mass enhancement in PrCo$_2$Ga$_8$. We note that this value of $\gamma$ is relatively high and is comparable to the value of 600~mJ/mol.K$^2$ found for the  heavy fermion superconductor PrOs$_4$Sb$_{12}$ \cite{bauer2002superconductivity}. 


%
\begin{figure}[!h]
	\centering
	\includegraphics[scale=2.0]{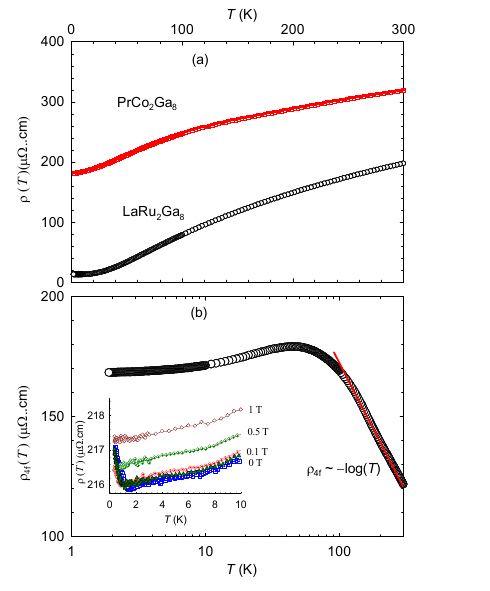}
	\caption{(a) The electrical resistivity of  PrCo$_2$Ga$_8$ and  LaRu$_2$Ga$_8$ in the  temperature range of 2 to 300~K. (b) A semi-log plot of electronic contribution to electrical resistivity, $\rho_{4f}(T)$ where the red solid line indicates a $-$log($T$) dependence of $\rho_{4f}(T)$. The inset is the field dependence of the electrical resistivity in the temperature range of 0.35 to 10 ~K. \label{fig2_rho}}
\end{figure}
\section{Electrical transport}
In Fig.~\ref{fig2_rho} (a) the electrical resistivity, $\rho(T)$ of  PrCo$_2$Ga$_8$ and  LaRu$_2$Ga$_8$ in the  temperature range of 2~K to 300~K are presented. The electrical resistivity $\rho(T)$ of  LaRu$_2$Ga$_8$ follows a typical metallic behaviour from 300~K down to about 2~K. The $\rho(T)$ of  PrCo$_2$Ga$_8$ shows a weak temperature dependence of the electrical resistivity down to about 100~K below which it develops a small change in slope down to low temperatures thereby losing about 20~\% of its resistivity between 100~K and 2~K. The residual resistivity ratio (RRR) estimated for PrCo$_2$Ga$_8$ is about 1.63 and is smaller compared to a value of 12.5 obtained for LaRu$_2$Ga$_8$. In Fig.~\ref{fig2_rho}(b), the electronic contribution to the electrical resistivity, $\rho_\mathrm{4f}(T)$  obtained by subtracting the resistivity of  LaRu$_2$Ga$_8$  from that of  PrCo$_2$Ga$_8$ is presented in the main panel. The $\rho_\mathrm{4f}(T)$ evolves in a semiconducting manner by following a gradual increase from 300~K down to about 60~K. Within this temperature ($T$) range, a $-$log($T$) dependence of $\rho_\mathrm{4f}(T)$ is implied. This behaviour is indicated by a $-$log($T$) dependence of $\rho_\mathrm{4f}(T)$ as shown by the fit (red solid line) in the main panel of Fig.~\ref{fig2_rho} (b). Below 60~K, the $\rho_\mathrm{4f}(T)$~ then shows a pronounced curvature before saturating towards lower temperatures. In iso-structural compound PrFe$_2$Ga$_8$ \cite{Ogunbunmi2017magnetization}, similar behaviour of   $-$ log($T$) dependence of $\rho_\mathrm{4f}(T) $  together with some pronounced Kondo-like features have been observed.\\
The field dependence of electrical resistivity of PrCo$_2$Ga$_8$ between 0.35~K and 10~K are presented in the inset of Fig.~\ref{fig2_rho}(b). A transition at $T_m$ = 1.28~K is observed in the 0~T result in support of the observation in the specific heat data. Below $T_m$, $\rho(T)$ shows an upturn and increases on further cooling indicating the formation of superzone gap \cite{das2012anisotropic}. Similar behaviour in electrical transport has been observed in other related quasi-skutterudite compounds \cite{Ogunbunmi2017magnetization,muro2011magnetic}. $T_m$ is tuneable in fields and it shifts to lower temperatures in applied fields up to 0.1~T. With a field of 0.5~T,  $T_m$ is completely suppressed and a positive magnetoresistance is observed with higher fields. \\
%
\begin{figure}[!t]
	\centering
	\includegraphics[scale=0.57]{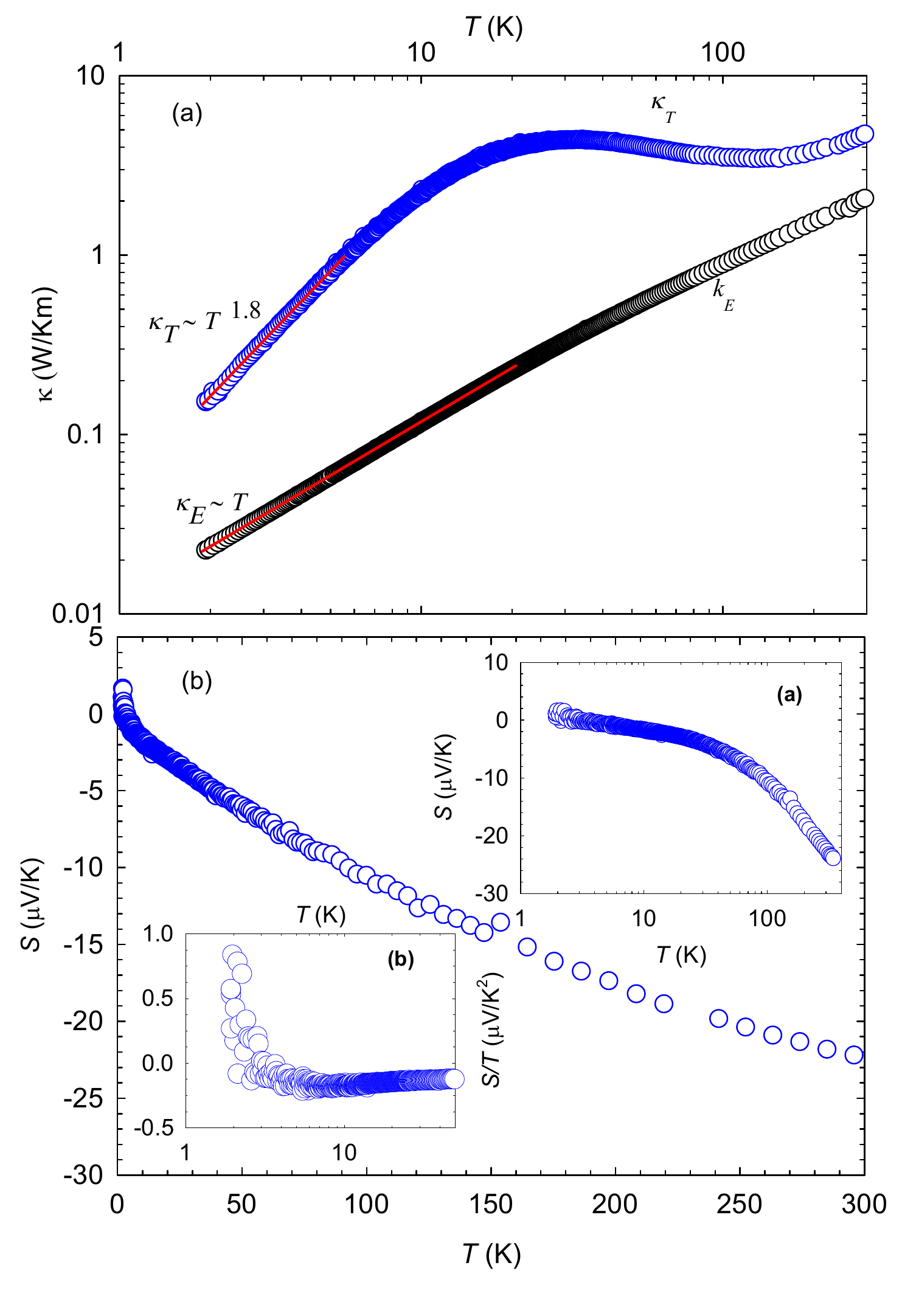}
	\caption{(a) The total thermal conductivity ($\kappa_T$) along with the electronic contribution ($\kappa_E$) are shown on a log-log scale. The red solid line shows the temperature dependence of $\kappa_T$ and $\kappa_E$ at low temperatures. (b) The thermopower variation with temperature in the range of 2 and 300~K is shown in the main panel. The inset (a) is the thermopower in a semi-log scale and inset (b) is a plot of $S(T)/T$ against $T$. \label{fig3_tto}}
\end{figure}
\section{Thermal transport}
The thermal conductivity, $\kappa_T$  of  PrCo$_2$Ga$_8$ is presented in Fig. \ref{fig3_tto}  (a) on a log-log scale. At room temperature the value of $\kappa_T$ is about 5.6 W/K.m which is relatively lower than the values found in ordinary metals. From room temperature down to about 50~K,  $\kappa_T$ shows a weak temperature dependence below which it shows a broad maximum centred around 30~K. Below this temperature, it then follows a power law dependence down to 2~K. Below 6~K,  $\kappa_T$ $\propto$ $T^{1.8}$ as  indicated by the red solid line. The electronic contribution to the thermal conductivity, $\kappa_E$ is deconvoluted from $\kappa_T$ by using the Wiedemann-Franz relation \cite{kittel2005introduction}, for the thermal conductivity in a highly degenerate electron gas: $\kappa_E$ = L$_0T$/$\rho(T)$,  where,  L$_0$ is the Lorentz number given as $\pi^2$k$_B^2$/3e$^2$  = 2.45 $\times$ 10$^{-8}$~ W.$\Omega$/K$^2$. The plot of $\kappa_E$ is also shown in Fig.~\ref{fig3_tto} (a). The temperature range between 2~K and 20~K can be described as $\kappa_E$ $\propto$ $T$ as shown by the red solid line in the low temperature region. This observation in electronic thermal conductivity agrees well with the behaviour of electronic thermal conductivity of a Fermi gas at low temperatures \cite{kittel2005introduction,falkowski2015cooperative}.  In the low temperature region, $\kappa_T$/$\kappa_E$ $\le$ 10 which suggest that the electrons are important heat carriers in the temperature range studied and participate in scattering mechanism in the system. \\
Not withstanding the assumptions under which we have calculated $\kappa_E$, it is interesting to note the aposite behaviour of electrons between electric and thermal conductivities: between 2 K and 20 K the total electrical resistivity increases by $\simeq 2\%$ (Fig.~\ref{fig2_rho}(a)), whereas in thermal transport the electronic thermal conductivity increases by as much as $\sim 10$ times (Fig.~\ref{fig3_tto}(a)) over the same temperature range.\\
The thermopower of PrCo$_2$Ga$_8$ measured between 2 and 300~K is presented in Fig. \ref{fig3_tto} (b). The thermopower is negative almost throughout the temperature range and shows a fairly linear temperature dependence. In the semi-log plot shown in the inset (a), a change in slope of thermopower can be seen below about 80~K with a weak temperature dependence. Since the thermopower is nearly negative throughout the temperature range considered, this gives an indication that electrons are the dominant charge carriers that provide the thermal voltage in the system. 
It has been shown \cite{behnia2004thermoelectricity} that the ratio of thermopower, $S(T)$ to $T$ and $\gamma$ form a constant value given by q = $S$N$_{Av}|e|/\gamma T$, where N$_{Av} |e|$ = 9.65 $\times$ 10$^4$~C/mol is the Faraday number and q is expected to have values close to $\pm$1 depending on the dominant carriers. A plot of $S(T)/T$ is presented in inset (b) of Fig.~\ref{fig3_tto}(b). Below about 4~K the thermopower shows an upturn and enhancement to higher values on further cooling in a manner similar to what was observed in $C_p(T)/T$. The zero temperature limit of $S(T)/T$ has been estimated by a linear extrapolation of the thermopower to temperatures below 2~K. Hence $|S(T)/T|_{T\rightarrow0}$ $\approx$ 7.52~$\mu$V/K$^2$ which is close to the value of 10 $\mu$V/K$^2$ ($\gamma$ $\approx$1.1~J/mol K$^2$) obtained for PrTa$_2$Al$_{20}$  and less than a value of 22~$\mu$V/K$^2$ ($\gamma$ $\approx$1~J/mol K$^2$) in PrV$_2$Al$_{20}$ \cite{sakai2011kondo,machida2015anomalous}. Taking $\gamma$ = 637~mJ/mol K$^2$, the calculated value of q = 1.13 is obtained which is close to the predicted value of 1. We note that this value is about three order of magnitude higher than the value of 0.003 obtained for PrTi$_2$Zn$_{20}$ and comparable to a value of about 1 found for PrTi$_2$Al$_{20}$ \cite{machida2015anomalous}. 

%
%

%
%

%
%

\section{Summary and Conclusion}
The quasi-skutterudite compound PrCo$_2$Ga$_8$ has been investigated through dc magnetic susceptibility, magnetization, specific heat, electrical resistivity, thermal conductivity and thermopower measurements. At variance to a singlet ground state predicted for Pr moment in this compound based on its local site symmetry of $C_s$ in the crystal structure, PrCo$_2$Ga$_8$ shows a phase transition below $T_m$ = 1.28~K. This transition is believed to have arisen due to dominant antiferromagnetic interactions in the system as suggested from the analysis of the magnetic susceptibility results. In previous studies, similar magnetic ground state has been observed for systems predicted to have a CEF singlet ground state. Typical examples are the iso-structural compounds PrCo$_2$Al$_8$ \cite{tougait2005prco} and PrFe$_2$Al$_8$ \cite{nair2016magnetic,nair2017pr}. Other relevant examples are PrCoAl$_4$ 	\cite{schobinger2001magnetic}, PrSi \cite{snyman2012anomalous} and PrTl$_3$  \cite{andres1972induced}. In these systems, the observation of magnetic order is largely attributed to the admixture of the singlet ground state with the CEF excited states under favourable exchange interaction value between the ground state singlet and the excited CEF levels. In this study, we have also observed the formation of a superzone gap as indicated by the upturn in the electrical resistivity below $T_m$.  The electrical resistivity show key signatures of Kondo-like behaviour in addition to possible heavy fermion signature observed in the specific heat analysis which gives a $\gamma$ = 637~mJ/mol.K$^2$. These observations made PrCo$_2$Ga$_8$ and other iso-structural gallides interesting systems to study. Further low temperature and neutron measurements are in progress to support the findings in this paper.
\section*{Acknowledgements}
MOO acknowledges the UJ-URC bursary. AMS thanks the SA-NRF (93549) and UJ-URC for financial support.

\section*{References}
\bibliography{PrCo2Ga8}
\bibliographystyle{elsarticle-num-names}

\end{document}